\documentstyle[twoside,epsfig]{article}

\catcode`\@=11
\long\def\@makefntext#1{
\protect\noindent \hbox to 3.2pt {\hskip-.9pt  
$^{{\eightrm\@thefnmark}}$\hfil}#1\hfill}		

\def\@makefnmark{\hbox to 0pt{$^{\@thefnmark}$\hss}}	
	
\def\ps@myheadings{\let\@mkboth\@gobbletwo
\def\@oddhead{\hbox{}
\rightmark\hfil\eightrm\thepage}   
\def\@oddfoot{}\def\@evenhead{\eightrm\thepage\hfil
\leftmark\hbox{}}\def\@evenfoot{}
\def\sectionmark##1{}\def\subsectionmark##1{}}

\oddsidemargin=\evensidemargin
\newcounter{sectionc}\newcounter{subsectionc}\newcounter{subsubsectionc}
\renewcommand{\section}[1] {\vspace{12pt}\addtocounter{sectionc}{1} 
\setcounter{subsectionc}{0}\setcounter{subsubsectionc}{0}\noindent 
	{\tenbf\thesectionc. #1}\par\vspace{5pt}}
\renewcommand{\subsection}[1] {\vspace{12pt}\addtocounter{subsectionc}{1} 
	\setcounter{subsubsectionc}{0}\noindent 
	{\bf\thesectionc.\thesubsectionc. {\kern1pt \bfit #1}}\par\vspace{5pt}}
\renewcommand{\subsubsection}[1] {\vspace{12pt}\addtocounter{subsubsectionc}{1}
	\noindent{\tenrm\thesectionc.\thesubsectionc.\thesubsubsectionc.
	{\kern1pt \tenit #1}}\par\vspace{5pt}}
\newcommand{\nonumsection}[1] {\vspace{12pt}\noindent{\tenbf #1}
	\par\vspace{5pt}}

\newcounter{appendixc}
\newcounter{subappendixc}[appendixc]
\newcounter{subsubappendixc}[subappendixc]
\renewcommand{\thesubappendixc}{\Alph{appendixc}.\arabic{subappendixc}}
\renewcommand{\thesubsubappendixc}
	{\Alph{appendixc}.\arabic{subappendixc}.\arabic{subsubappendixc}}

\renewcommand{\appendix}[1] {\vspace{12pt}
        \refstepcounter{appendixc}
        \setcounter{figure}{0}
        \setcounter{table}{0}
        \setcounter{lemma}{0}
        \setcounter{theorem}{0}
        \setcounter{corollary}{0}
        \setcounter{definition}{0}
        \setcounter{equation}{0}
        \renewcommand{\thefigure}{\Alph{appendixc}.\arabic{figure}}
        \renewcommand{\thetable}{\Alph{appendixc}.\arabic{table}}
        \renewcommand{\theappendixc}{\Alph{appendixc}}
        \renewcommand{\thelemma}{\Alph{appendixc}.\arabic{lemma}}
        \renewcommand{\thetheorem}{\Alph{appendixc}.\arabic{theorem}}
        \renewcommand{\thedefinition}{\Alph{appendixc}.\arabic{definition}}
        \renewcommand{\thecorollary}{\Alph{appendixc}.\arabic{corollary}}
        \renewcommand{\theequation}{\Alph{appendixc}.\arabic{equation}}
        \noindent{\tenbf Appendix \theappendixc #1}\par\vspace{5pt}}
\newcommand{\subappendix}[1] {\vspace{12pt}
        \refstepcounter{subappendixc}
        \noindent{\bf Appendix \thesubappendixc. {\kern1pt \bfit #1}}
	\par\vspace{5pt}}
\newcommand{\subsubappendix}[1] {\vspace{12pt}
        \refstepcounter{subsubappendixc}
        \noindent{\rm Appendix \thesubsubappendixc. {\kern1pt \tenit #1}}
	\par\vspace{5pt}}

\newcommand{\textlineskip}{\baselineskip=13pt}
\newcommand{\smalllineskip}{\baselineskip=10pt}

\def\eightcirc{
\begin{picture}(0,0)
\put(4.4,1.8){\circle{6.5}}
\end{picture}}
\def\eightcopyright{\eightcirc\kern2.7pt\hbox{\eightrm c}} 

\newcommand{\copyrightheading}[1]
	{\vspace*{-2.5cm}\smalllineskip{\flushleft
	{\footnotesize International Journal of Modern Physics A, #1}\\
	{\footnotesize $\eightcopyright$\, World Scientific Publishing
	 Company}\\
	 }}

\def\abstracts#1#2#3{{
	\centering{\begin{minipage}{4.5in}\baselineskip=10pt\footnotesize
	\parindent=0pt #1\par 
	\parindent=15pt #2\par
	\parindent=15pt #3
	\end{minipage}}\par}} 

\def\keywords#1{{
	\centering{\begin{minipage}{4.5in}\baselineskip=10pt\footnotesize
	{\footnotesize\it Keywords}\/: #1
	 \end{minipage}}\par}}

\renewenvironment{thebibliography}[1]
	{\frenchspacing
	 \ninerm\baselineskip=11pt
	 \begin{list}{\arabic{enumi}.}
	{\usecounter{enumi}\setlength{\parsep}{0pt}
	 \setlength{\leftmargin 12.7pt}{\rightmargin 0pt} 
	 \setlength{\itemsep}{0pt} \settowidth
	{\labelwidth}{#1.}\sloppy}}{\end{list}}

\newcounter{itemlistc}
\newcounter{romanlistc}
\newcounter{alphlistc}
\newcounter{arabiclistc}

\newcommand{\fcaption}[1]{
        \refstepcounter{figure}
        \setbox\@tempboxa = \hbox{\footnotesize Fig.~\thefigure. #1}
        \ifdim \wd\@tempboxa > 5in
           {\begin{center}
        \parbox{5in}{\footnotesize\smalllineskip Fig.~\thefigure. #1}
            \end{center}}
        \else
             {\begin{center}
             {\footnotesize Fig.~\thefigure. #1}
              \end{center}}
        \fi}

\newcommand{\tcaption}[1]{
        \refstepcounter{table}
        \setbox\@tempboxa = \hbox{\footnotesize Table~\thetable. #1}
        \ifdim \wd\@tempboxa > 5in
           {\begin{center}
        \parbox{5in}{\footnotesize\smalllineskip Table~\thetable. #1}
            \end{center}}
        \else
             {\begin{center}
             {\footnotesize Table~\thetable. #1}
              \end{center}}
        \fi}

\def\@citex[#1]#2{\if@filesw\immediate\write\@auxout
	{\string\citation{#2}}\fi
\def\@citea{}\@cite{\@for\@citeb:=#2\do
	{\@citea\def\@citea{,}\@ifundefined
	{b@\@citeb}{{\bf ?}\@warning
	{Citation `\@citeb' on page \thepage \space undefined}}
	{\csname b@\@citeb\endcsname}}}{#1}}

\newif\if@cghi
\def\cite{\@cghitrue\@ifnextchar [{\@tempswatrue
	\@citex}{\@tempswafalse\@citex[]}}
\def\citelow{\@cghifalse\@ifnextchar [{\@tempswatrue
	\@citex}{\@tempswafalse\@citex[]}}
\def\@cite#1#2{{$\null^{#1}$\if@tempswa\typeout
	{IJCGA warning: optional citation argument 
	ignored: `#2'} \fi}}

\def\pmb#1{\setbox0=\hbox{#1}
	\kern-.025em\copy0\kern-\wd0
	\kern.05em\copy0\kern-\wd0
	\kern-.025em\raise.0433em\box0}

\def\fnt#1#2{\footnotetext{\kern-.3em
	{$^{\mbox{\scriptsize #1}}$}{#2}}}

\def\fpage#1{\begingroup
\voffset=.3in
\thispagestyle{empty}\begin{table}[b]\centerline{\footnotesize #1}
	\end{table}\endgroup}

\def\runninghead#1#2{\pagestyle{myheadings}
\markboth{{\protect\footnotesize\it{\quad #1}}\hfill}
{\hfill{\protect\footnotesize\it{#2\quad}}}}
\font\tenrm=cmr10
\font\tenit=cmti10 
\font\tenbf=cmbx10
\font\bfit=cmbxti10 at 10pt
\font\ninerm=cmr9

\font\eightrm=cmr8

\def\qed{\hbox{${\vcenter{\vbox{			
   \hrule height 0.4pt\hbox{\vrule width 0.4pt height 6pt
   \kern5pt\vrule width 0.4pt}\hrule height 0.4pt}}}$}}

\newcommand{\case}[2]{\mbox{\footnotesize $\displaystyle \frac{#1}{#2}$}}

\begin{document}

\runninghead{Contemporary Applications of 
DSEs} {Contemporary Applications of 
DSEs}

\normalsize\textlineskip
\thispagestyle{empty}
\setcounter{page}{1}

\copyrightheading{}			

\vspace*{0.88truein}

\fpage{1}
\centerline{\bf DYSON-SCHWINGER EQUATIONS}
\vspace*{0.035truein}
\centerline{\bf -- ASPECTS OF THE PION}
\vspace*{0.37truein}
\centerline{\footnotesize M.B.~HECHT, C.D.~ROBERTS and S.M.~SCHMIDT}
\vspace*{0.015truein}
\centerline{\footnotesize\it Physics Division, Bldg 203, Argonne National
Laboratory}
\baselineskip=10pt
\centerline{\footnotesize\it Argonne, IL 60439-4843, USA}

\vspace*{0.21truein}
\abstracts{
\centerline{\parbox{30em}{\sc Contribution to the Proceedings of
``DPF$\,$2000,'' the Meeting of the Division of Particles and Fields
of the American Physical Society, August 9-12, 2000, Department of
Physics, the Ohio State University, Columbus, Ohio.}\\[2ex]}
The contemporary use of Dyson-Schwinger equations in hadronic physics
is exemplified via applications to the calculation of pseudoscalar
meson masses, and inclusive deep inelastic scattering with a
determination of the pion's valence-quark distribution function.
}{}{}
\vspace*{1ex}

\keywords{
Dyson-Schwinger equations, 
Goldstone bosons,
Heavy-Quarks,
Valence-quark distribution functions.
}

\textlineskip			
\vspace*{12pt}			

\vspace*{1pt}\textlineskip	

\noindent
The Dyson-Schwinger equations (DSEs)\cite{cdragw} provide an approach
well-suited to the calculation of pion observables.  Since a chiral
symmetry preserving truncation scheme exists,\cite{truncscheme} they
provide a framework in which the dichotomous
bound-state/Goldstone-mode character of the pion is easily
captured.\cite{mrt98,mr97} Furthermore, because perturbation theory is
recovered in the weak coupling limit, they combine; e.g., a
description of low-energy $\pi$-$\pi$ scattering\cite{pipi} with a
calculation of the electromagnetic pion form factor, $F_\pi(q^2)$,
that yields\cite{mrpion} the $1/q^2$-behaviour expected from
perturbative analyses at large spacelike-$q^2$ and a calculated
evolution to the $\rho$-meson pole in the timelike
region.\cite{mtpion} These and other contemporary applications are
reviewed in Refs.\ [\ref{revbastiR},\ref{revreinhardR}].

As an illustration, using the inhomogeneous Bethe-Salpeter equations for the
axial-vector and pseudovector vertices; the dressed-quark DSE; and the fact
that a nonperturbative Ward-Takahashi identity preserving truncation of the
DSEs is possible, it was shown in Ref.\ [\ref{mrt98R}] that for flavour
nonsinglet pseudoscalar mesons
\begin{equation}
\label{gmor}
\displaystyle f_H m_H^2 = {\cal M}_H^\zeta r_H^\zeta\,,
\end{equation}
with ${\cal \cal M}_H^\zeta:= {\rm tr}_{\rm flavour}
[M_{(\zeta)}\,\{T^H,(T^H)^{\rm t}\}]$\,, where $M_{(\zeta)}={\rm
diag}(m_u^\zeta,m_d^\zeta,m_s^\zeta,\ldots)$, with $\zeta$ the
renormalisation point, and $(\cdot)^{\rm t}$ indicates matrix
transpose, so that ${\cal \cal M}_H^\zeta$ is the sum of the
constituents' current-quark masses.  This model-independent identity
is valid for all current-quark masses, irrespective of their
magnitude, and therefore provides a single formula that unifies the
light- and heavy-quark regimes.

In Eq.\ (\ref{gmor}), $f_H$ is the leptonic decay constant,
\begin{equation}
\label{fH}
f_H\, P_\mu = Z_2\int^\Lambda\!\case{d^4 q}{(2\pi)^4}\,\case{1}{2} {\rm
tr}\left[\left(T^H\right)^{\rm t} \gamma_5 \gamma_\mu \chi_H(q;P)\right]\,,
\end{equation}
where $Z_2=Z_2(\zeta,\Lambda)$ is the dressed-quark wave function
renormalisation constant, with $\Lambda$ the regularisation
mass-scale.  (The advantages of employing a translationally invariant
regularisation scheme when Ward-Takahashi identities are involved
should be obvious.)  The r.h.s.\ is \underline{gauge-invariant} and
independent of $\zeta$ and $\Lambda$.  The other factor is
\begin{equation}
\label{rH}
i r_H^\zeta\, = Z_4\int^\Lambda_q\,\case{1}{2} {\rm
tr}\left[\left(T^H\right)^{\rm t} \gamma_5 \chi_H(q;P)\right]\,,
\end{equation}
where $Z_4=Z_4(\zeta,\Lambda)$ is the dressed-quark mass
renormalisation constant.  Here the r.h.s.\ is gauge invariant and
cutoff-independent, and $Z_4$'s $\zeta$-dependence ensures that the
product on the r.h.s.\ of Eq.\ (\ref{gmor}) is independent of the
renormalisation point.  Using these formulae it can be established
that Eq.\ (\ref{gmor}) reproduces the so-called
Gell-Mann--Oakes-Renner relation in the limit of small current-quark
masses\cite{mrt98} and also predicts that heavy-meson masses increase
linearly with the mass of their heaviest constituent.\cite{mishasvy}
However, the latter ``heavy-quark limit'' provides a poor
approximation to the pseudoscalar meson mass trajectory for
current-quark masses less than that of the $b$-quark.\cite{qciv}

\begin{figure}[t]
\centerline{\epsfig{figure=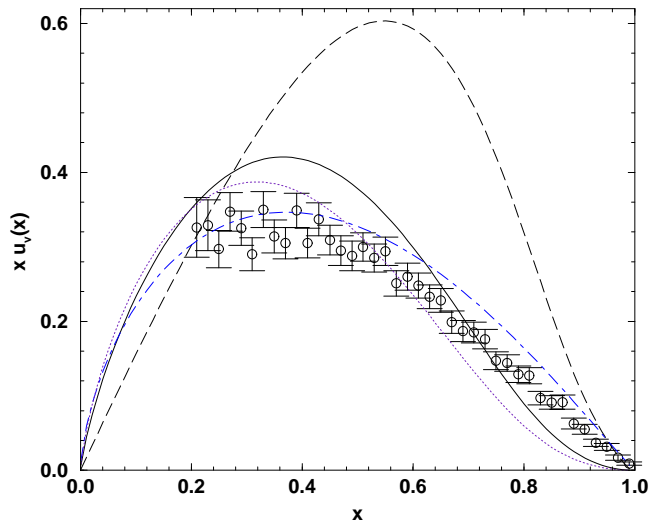,height=7.0cm}}
\caption{\label{figure} Dashed line: calculated form of $x
u_v(x;q_0)$; Solid line: evolved distribution, $x u_v(x;q=2\,{\rm
GeV})$; Dotted line: $x u_v(x;q=4.05\,{\rm GeV})$, evolved from $x
u_v(x;q=2\,{\rm GeV})$ with $\Lambda^{\rm N_f=4}_{\rm QCD}=
0.204\,$GeV; and Dot-dashed line: phenomenological fit in Ref.\
[\protect\ref{suttonR}], which takes the form $xu_v(x)\propto x^\alpha
(1-x)^\beta$.  The data\protect\cite{DYexp2} are obtained with an
invariant $\mu^+ \mu^-$-mass $>$ $4.05\,$GeV and inferred from the
differential pion-nucleon Drell-Yan cross section using simple
distribution parametrisations of the type just indicated, yielding
$\alpha=0.64\pm0.03$, $\beta=1.15\pm0.02$.  This data was part of the
set employed in the fit of Ref.\ [\protect\ref{suttonR}].  Fits to our
calculation using the parametrisation in Eq.\ (\protect\ref{fit94})
are indistinguishable from our result on the scale of this figure.}
\end{figure}

Another more recent application is a calculation of the pion's
valence-quark distribution,\cite{piux} of which a very accurate
measurement is possible given a high-luminosity electron-proton
collider.\cite{roypaul} Using an algebraic DSE-model,\cite{mark}
employed successfully in studies of a wide range of hadronic
observables; e.g., Refs.\ [\ref{mishasvyR},\ref{qcivR}], it is
straightforward to calculate the ``handbag'' contributions to the
virtual photon-pion forward Compton scattering amplitude.  These are
the only impulse approximation diagrams that survive when calculating
the pion's structure function in the deep inelastic Bjorken limit and
yield the results in Fig.\ \ref{figure}.

In this calculation,$\!\!$\footnote{Herein we have corrected a minor
numerical error discovered in the calculations of Ref.\
[\protect\ref{piuxR}].}
$\,$dressed-quarks with a valence-quark mass of $\check M=301\,$MeV
carry $71\,$\% of the pion's momentum at a resolving scale
$q_0=0.54\,$GeV$\,=1/(0.37\,{\rm fm})$.  The remainder is carried by
the gluons that effect the binding of the pion bound state.  The
second and third moments of the distribution are $\langle
x^2\rangle_{q_0} =0.18$, $\langle x^3\rangle_{q_0}=0.10$.  To
determine the resolving scale, $q_0$, we employed the $3$-flavour
($\Lambda_{\rm QCD}^{N_f=3}=0.242\,$GeV), leading-order, nonsinglet
renormalisation group (evolution) equations to evolve the distribution
in Fig.\ \ref{figure} up to $q=2\,$GeV, and required agreement between
the first and second moments of our evolved distribution and those
calculated from the phenomenological fits in Ref.\ [\ref{suttonR}].
$q_0=0.54\,$GeV gives
\begin{equation}
\begin{array}{l|lll}
                &       \langle x\rangle_q   & \langle x^2 \rangle_q & 
        \langle x^3 \rangle_q \\\hline
{\rm Calc.}\mbox{\cite{piux}}     &        0.24                   & 0.098 & 0.049      \\
{\rm Fit}\mbox{\cite{sutton}}& 0.24\pm 0.01    & 0.10\pm 0.01 &  
                0.058 \pm 0.004      \\
{\rm Latt.}\mbox{\cite{lattice}}
                & 0.27 \pm 0.01                 & 0.11 \pm 0.3 & 
                0.048 \pm 0.020 
\end{array}
\end{equation}
with the valence-quarks now carrying a momentum-fraction of $0.49$.  

The evolved distribution is also shown in Fig.\ \ref{figure}.  The
evident accentuation via evolution of the convex-up behaviour of the
distribution near $x= 1$ is characteristic of the renormalisation
group equations, which populate the sea-quark distribution at
small-$x$ at the expense of large-$x$ valence-quarks.  The simple
parametrisation:\cite{sutton} $x\,u_v(x;q) \propto x^\alpha \,
(1-x)^\beta$, is not flexible enough to provide a good pointwise fit
to our calculated distribution.  However, the modernised fitting
form\cite{mrs94}
\begin{equation}
\label{fit94}
x\, u_v(x) \propto
x^{\eta_1}\,(1-x)^{\eta_2}\,(1-\epsilon_u\,\sqrt{x}+\gamma_u \,x)\,,
\end{equation}
with parameters: $\eta_1$, $\eta_2$, $\epsilon_u$, $\gamma_u$, can
describe our calculated result very well.

The importance of this is apparent when one appreciates that the
functional form $x\,u_v(x;q) \propto x^\alpha \, (1-x)^\beta$,
$\alpha=0.67$, $\beta=1.13$, can be obtained\cite{piux,arriola} via
the evolution from $q_0=0.35\,$GeV of $u_v(x)=\theta(x)\theta(1-x)$,
which latter distribution corresponds to the valence-quark carrying
each and every fraction of the pion's momentum with equal probability,
and to a {\it momentum-independent} pion Bethe-Salpeter amplitude;
i.e., it is equivalent to representing the pion as a point particle.
The convex-up character of our result is a characteristic feature of
calculations in which the pion is described by a finite-size
Bethe-Salpeter amplitude (see, e.g., Refs.\ [\ref{dorokhovR}]).  Hence
the convexity of the valence-quark distribution can be interpreted as
a signature of binding in a nonpointlike pion.

While the moments of our calculated distribution are indistinguishable
from those of that fitted to data, there is a pointwise discrepancy
between our result and the data.  Currently we judge that this
discrepancy can be attributed to the restricted function space used
thus far in parametrising pion data, especially as the modernised
fitting form, Eq.\ (\ref{fit94}), can describe our calculation.  We
would be much interested in a reanalysis of existing data using the
updated parametrisations and, indeed, in new experiments; e.g, Ref.\
[\ref{roypaulR}], with small errors on the valence-quark domain, which
might expose what we have described as the signature of
quark-antiquark binding in QCD's Goldstone mode.

\vspace*{1ex}

\noindent
{\bf Acknowledgements.}~~We acknowledge useful communications with
J.C.R.~Bloch.  This work was supported by the US Department of Energy,
Nuclear Physics Division, under contract no. W-31-109-ENG-38 and benefited
from the resources of the National Energy Research Scientific Computing
Center.  SMS is grateful for financial support from the A.\ v.\ Humboldt
foundation.

\nonumsection{References}

\end{document}